\documentstyle[11pt,newpasp,twoside,epsf]{article}
\begin{document}
\title{Self-regulated hydrodynamical process
  in halo stars : a possible explanation of the lithium plateau}
\author{Sylvie Th\'eado and Sylvie Vauclair}
\affil{Laboratoire d'Astrophysique, 14 av. Ed. Belin, 31400 Toulouse,
  France}

\begin{abstract}
It has been known for a long time (Mestel~1953) that the meridional
circulation velocity in stars, in the presence of 
$\displaystyle \mu $-gradients, is the sum of two terms, one due to
the classical thermal imbalance ($\displaystyle \Omega$-currents)
and the other one due to the induced horizontal
$\displaystyle \mu $-gradients ($\displaystyle \mu $-induced
currents, or $\displaystyle \mu $-currents in short). In the most
general cases, $\displaystyle \mu $-currents are opposite to
$\displaystyle \Omega$-currents.  Vauclair (1999) has shown that
such processes can, in specific cases, lead to a quasi-equilibrium
stage in which both the circulation and the helium settling is frozen.
Here we present computations of the circulation currents in halo star
models, along the whole evolutionary sequences for four stellar masses
with a metallicity of [Fe/H] = -2. We show that such a self-regulated
process can account for the constancy of the lithium abundances and the
small dispersion in the Spite plateau.
\end{abstract}

>From spectroscopic observations, the lithium abundance in main
sequence Pop~II field stars with effective temperatures larger than
5500~K is remarkably constant, with a very low dispersion if any (Spite
\& Spite 1982;  Spite et al. 1996; Bonifacio \& Molaro 1997; Molaro 1999), while 
large lithium abundance dispersions do occur for Pop I
stars.
We claim that the reason for this behavior may be due to the
self-regulating process in slowly rotating stars as described by
Vauclair (1999).

In rotating stars, the equipotentials of  ``effective gravity''
(including the centrifugal acceleration) have
ellipsoidal shapes while the energy transport still occurs in a
spherically
symetrical way. The resulting thermal imbalance must be compensated
by macroscopic motions: the so-called ``meridional circulation''(Von Zeipel 1924;
Mestel 1953;
Maeder \& Zahn 1998). 

In the presence of vertical $\displaystyle \mu $-gradients,
the circulation velocity is the sum of two terms, one which does not
depend on $\displaystyle \mu $ (the so-called ``$\displaystyle \Omega
$ currents'') and one which gathers the $\displaystyle \mu $
dependent terms
(the ``$\displaystyle \mu $ currents'').

In the present paper, we have computed the $\displaystyle
\Omega$-currents and
the $\displaystyle \mu $-currents along the evolutionary track of a .75 solar mass halo
stars.
All the parameters included in the computations are the same as for the solar models
(Richard 1999). The horizontal $\displaystyle \mu $ gradients 
are derived using Zahn (1992) theory of anisotropic turbulence (see
Vauclair (1999 and 2000) for details).
The lithium variations with time are then computed within the same framework. 

\begin{figure}
\plotfiddle{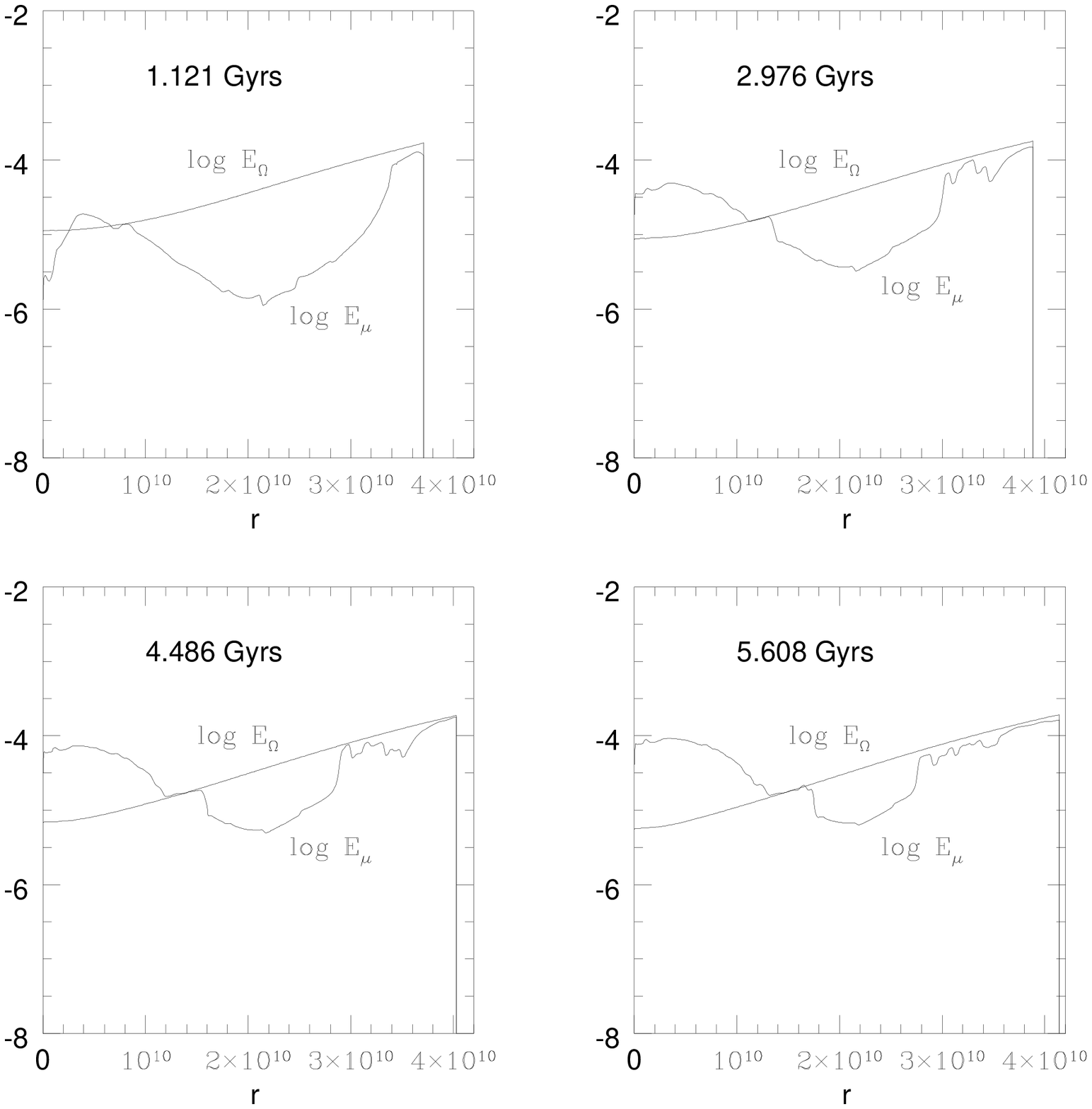}{6.0cm}{0}{40}{40}{-220}{-90}
\plotfiddle{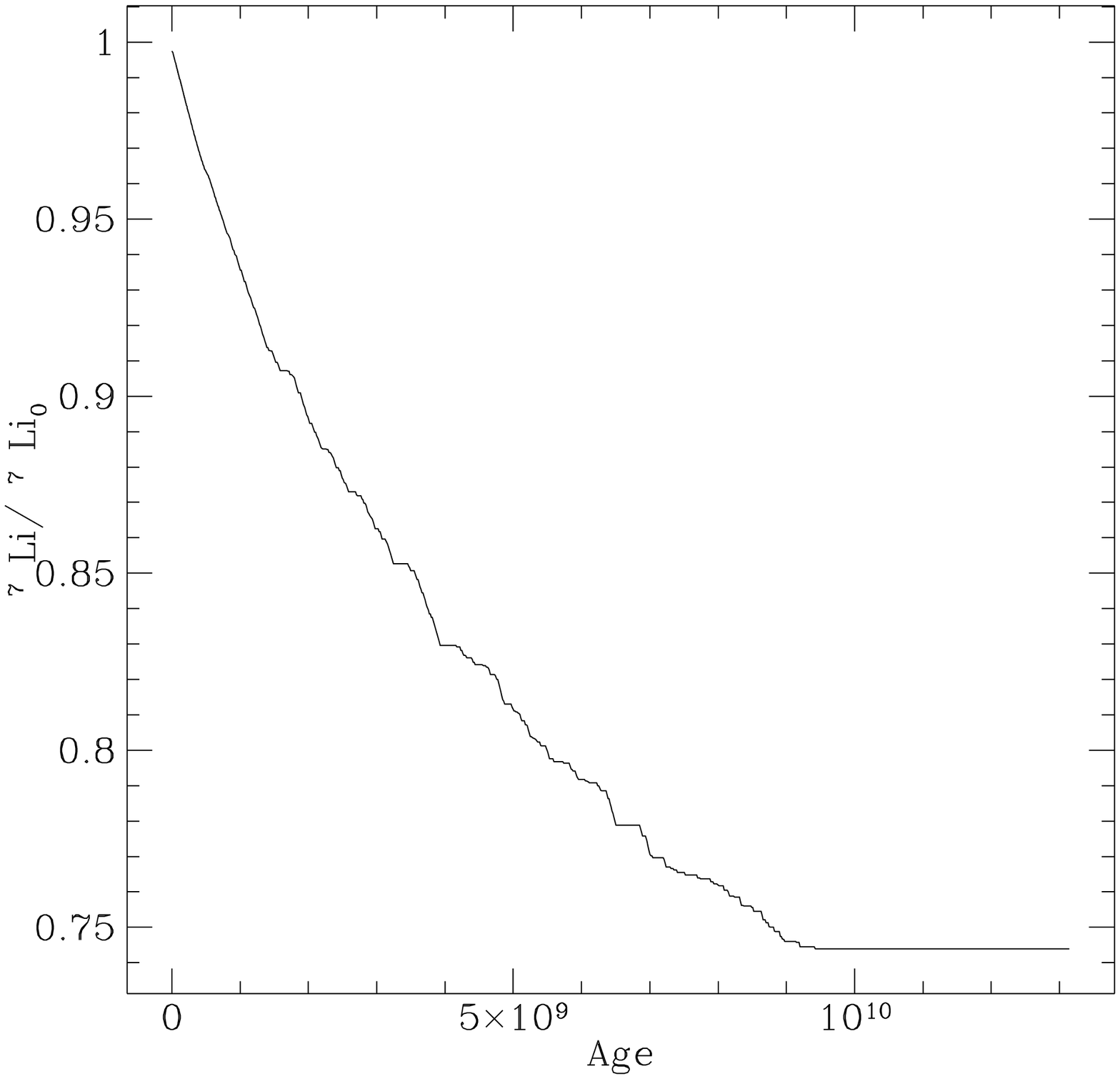}{0cm}{0}{30}{40}{20}{-65}
\caption{\small{Computations of the $\Omega$-currents and $\mu $-currents
 and lithium abundance variation with time 
in a $0.75 M_{\odot}$ halo stars with [Fe/H]=-2.}}
\end{figure}

The $\displaystyle \mu $-currents increase with time below the convective zone because of
helium settling (it also increases in the
core because of nuclear reactions).
An equilibrium situation soon occurs below the convection zone, for
which the two currents become equal. Then, the circulation freezes out
as well as the gravitational settling. Lithium decreases very slowly and remains constant
when the whole star
is ``frozen''. The depletion is not larger than $25\%$.
This can explain the very small dispersion observed in the Spite plateau.

There are many observations in stars which give evidences of mixing
processes
occuring below the outer convection zones as, for example, the lithium
depletion
observed in the Sun and in galactic clusters. The process we have
described
here should not apply in all these stars. The reason could be related to
the
rapid rotation of young stars on the ZAMS and to their subsequent
 rotational braking and differential rotation, which is not supposed
here to take place in halo stars.

\end{document}